\documentclass[12pt, a4paper]{article}
\usepackage[top=1in, bottom=1in, left=1.25in, right=1.25in]{geometry}
\usepackage{booktabs}
\usepackage{multirow}
\usepackage{setspace}
\usepackage{natbib}
\usepackage{authblk}
\usepackage{setspace}
\usepackage[pdftex]{graphicx}
\usepackage{amsmath}
\usepackage{amsthm}
\usepackage{amsfonts}
\usepackage{ascmac}
\usepackage{amssymb}
\usepackage{bm}
\usepackage{color}
\usepackage{enumitem}
\usepackage{type1cm} 
\usepackage{mathrsfs}
\usepackage{hyperref}
\usepackage{comment}

\newtheorem{theorem}{Theorem}

\usepackage{comment}
\usepackage{enumitem}
\usepackage{multicol}
\usepackage{multirow}
\usepackage{chngpage}
\usepackage{rotating}

\theoremstyle{plain}

\newcommand{\diag}{{\rm diag}}

\newcommand{\pd}{\partial}

\title{\bf Geometric Mean Type of Proportional Reduction in Variation Measure for Two-Way Contingency Tables}
\author[1]{Wataru Urasaki}
\author[1]{Yuki Wada}
\author[2]{Tomoyuki Nakagawa}
\author[1]{Kouji Tahata}
\author[2,3]{Sadao Tomizawa}

\affil[1]{Department of Information Sciences, Tokyo University of Science}
\affil[2]{School of Data Science, Meisei University}
\affil[3]{Department of Information Science, Meisei University}
\date{}

\begin{document}

\clearpage
\pagenumbering{arabic}
\maketitle

\begin{abstract}
In a two-way contingency table analysis with explanatory and response variables, the analyst is interested in the independence of the two variables.
However, if the test of independence does not show independence or clearly shows a relationship, the analyst is interested in the degree of their association.
Various measures have been proposed to calculate the degree of their association, one of which is the proportional reduction in variation (PRV) measure which describes the PRV from the marginal distribution to the conditional distribution of the response.
The conventional PRV measures can assess the association of the entire contingency table, but they can not accurately assess the association for each explanatory variable.
In this paper, we propose a geometric mean type of PRV (geoPRV) measure that aims to sensitively capture the association of each explanatory variable to the response variable by using a geometric mean, and it enables analysis without underestimation when there is partial bias in cells of the contingency table.
Furthermore, the geoPRV measure is constructed by using any functions that satisfy specific conditions, which has application advantages and makes it possible to express conventional PRV measures as geometric mean types in special cases.
\end{abstract}

\medskip


\medskip


\section{Introduction}\label{sec1}
Categorical variables are formed from categories and are employed in various fields such as medicine, psychology, education, and social science.
Considering two types of categorical variables, one consisting of $R$ categories and the other consisting of $C$ categories.
These two variables have $R \times C$ combinations, which can be represented in a table with $R$ rows and $C$ columns.
This is called a two-way contingency table, where each ($i, j$) cell ($i=1,2,\ldots, R;~j=1,2,\ldots, C$) displays only the observed frequencies.
Typically, the two-way contingency table is used to evaluate whether the two variables are related, i.e., statistically independent.
If the independence of the two variables is rejected for example by Pearson's chi-square test, or they are clearly considered to be related, we are interested in the strength of their association.

As a method to investigate the associative structure of the contingency table, association models have been proposed by \cite{gilula1986canonical}, \cite{goodman1981association, goodman1985analysis}, and \cite{rom1992generalized}. 
This method can determine whether there is a relationship between row and column variables by the goodness-of-fit test with models.
However, this method only focuses on whether or not there is a relationship, and we can not quantitatively determine what the degree of association is.

Instead of the goodness-of-fit test with the models, a variety of measures have been proposed as indicators that can show the degree of association within the interval from 0 to 1 by \cite{agresti2003categorical}, \cite{bishop2007discrete}, \cite{cramer1999mathematical},  \cite{everitt1992analysis},\cite{tomizawa2004generalization}, and \cite{Tschuprow1926grundbegriffe, Tschuprow1939principles}.
These measures calculate the degree of deviation from independence for each $(i, j)$ cell in the contingency table and derive the degree of association from the sum of all cells.
Because of the method, these measures can be applied to most contingency tables without distinguishing whether row and column variables are explanatory or response variables.
However, in actual contingency table analysis, there are cases where the row and column variables are defined as explanatory or response variables.
In such cases, it is not appropriate to analyze each variable by ignoring its characteristics.

Alternative measures have been proposed by \cite{doi:10.1080/01621459.1954.10501231}, and \cite{theil1970estimation}, which is explained by the proportional reduction in variation (PRV) from the marginal distribution to the conditional distributions of the response.
The measures constructed by the method is called PRV measure.
The PRV measure is an important tool in summarizing the strength of association of the entire contingency table because the way it is constructed makes it easy to interpret the values.
In addition, we sometimes want to focus on the association of some categories of explanatory variables, but conventional PRV measures underestimate the strength and thus may not be able to accurately reflect the partial association numerically.
In the study of models and scales for evaluating the symmetry of the contingency table, \cite{nakagawa2020geometric}, \cite{saigusa2016measure}, and \cite{saigusa2019measure} proposed to evaluate the partial symmetry by using the geometric mean.
On the other hand, little research has been done in the case of the partial association.

In this paper, we propose a geometric mean type of PRV (geoPRV) measure via a geometric mean and functions satisfying certain conditions.
Therefore, the geoPRV measure has application advantages and makes it possible to express previously proposed PRV measures as geometric mean types in special cases.
By using the geometric mean to sensitively capture the association of each explanatory variable, analysis can be performed without underestimating the degree of association when cells in the contingency table are partially biased. 
In addition, the geoPRV measure enables us to know local association structures.
Furthermore, the geoPRV measure can be analyzed regardless of whether the categorical variable is nominal or ordinal because its value does not change even when rows and columns are swapped.
The rest of this paper is organized as follows. 
Section 2 introduces previous research on an extension of generalized PRV (eGPRV) measure and proposes the geoPRV measure.
Section 3 presents the approximate confidence intervals of the proposed measures. 
Section 4 confirms the values and confidence intervals of the proposed measure using several artificial and actual data sets, and compares them with the eGPRV measure.
Section 5 presents our conclusions.

\section{PRV Measure}\label{sec2}
In this Section, we introduce measures using function $f(x)$ that satisfy the following conditions: (i) The function $f(x)$ is convex function; (ii) $0 \cdot f(0/0)=0$; (iii) $\lim_{x\to +0}f(x)=0$; (iv) $f(1)=0$.
Examples of the function are introduced, and models and measures using it have been proposed by \cite{kateri1994f}, \cite{momozaki2022extension} and \cite{tahata2022advances}.
These proposals are intended to generalize existing models and measures and have application advantages that make it easy to construct new ones and allow adjustments with tuning parameters to fit the analysis.
Section 2.1 provides some conventional PRV measures by \cite{momozaki2022extension}. 
In Section 2.2, we propose a geometric mean type of PRV measure and its characteristics.

\subsection{Conventional PRV Measure}
Consider $R\times C$ contingency table with nominal categories of the explanatory variable $X$ and the response variable $Y$.
Let $p_{ij}$ denote the probability that an observation will fall in the $i$th row and $j$th column of the table ($i=1,\ldots, R;j=1,\ldots, C)$.
In addition, $p_{i\cdot}$ and $p_{\cdot j}$ are denoted as $p_{i\cdot}=\sum_{l=1}^C p_{il}$, $p_{\cdot j}=\sum_{k=1}^Rp_{kj}$.
The conventional PRV measure has the form
\[
\Phi = 
\frac{V(Y)-E[V(Y \vert X)]}{V(Y)} = 
\frac{\displaystyle V(Y)-\sum_{i=1}^Rp_{i\cdot}V(Y \vert X=i)}{V(Y)},
\]
where $V(Y)$ is a measure of variation for the marginal distribution of $Y$, and $E[V(Y \vert X)]$ is the expectation for the conditional variation of $Y$ given the distribution of $X$ (see, \citealp{agresti2003categorical}).
$\Phi$ is using the weighted arithmetic mean of $V(Y \vert X=i)$, i.e, $\sum_{i=1}^Rp_{i\cdot}V(Y \vert X=i)$.
By changing the variation measure, various PRV measures can be expressed, such as {\it uncertainty coefficient} $U$ for the variation measure $V(Y)=-\sum_{j=1}^C p_{\cdot j}\log p_{\cdot j}$ called {\it Shannon entropy} and {\it concentration coefficient} $\tau$ for the variation measure $V(Y)=1-\sum_{j=1}^C p_{\cdot j}^2$ called {\it Gini concentration} (see, \citealp{agresti2003categorical}).
\cite{tomizawa1997generalized} proposed a generalized PRV measure $T^{(\lambda)}$ that includes $U$ and $\tau$ by using $V(Y) = \left( 1-\sum_{j=1}^C p_{\cdot j}^{\lambda+1} \right)/\lambda$ as the variation measure which is \cite{patil1982diversity} {\it diversity index of degree $\lambda$} for the marginal distribution $p_{\cdot j}$.
Furthermore, \cite{momozaki2022extension} proposed an extension of generalized PRV (eGPRV) measure that includes $U$, $\tau$, and $T^{(\lambda)}$:
\[
\Phi_f = \frac{\displaystyle-\sum_{j=1}^{C}f(p_{\cdot j}) - \sum_{i=1}^R p_{i\cdot} \left[ - \sum _{j=1} ^C f \left(p_{ij}/p_{i\cdot} \right) \right]}{\displaystyle- \sum_{j=1}^C f(p_{\cdot j})}.
\]
The variation measure used in the eGPRV measure $\Phi_f$ are $V(Y) = - \sum_{j=1}^{C}f(p_{\cdot j})$.

\subsection{Geometric Mean Type of PRV Measure}
We propose a new PRV measure by using the weighted geometric mean of $V(Y \vert X=i)$ that aims to sensitively capture the association of each explanatory variable to the response variable. 
Assume that $p_{\cdot j}>0$ and $V(Y \vert X=i)$ is a real number greater than or equal to 0 ($i=1,\ldots, R;~j=1,\ldots, C$).
We propose a geometric mean type of PRV (geoPRV) measure for $R\times C$ contingency tables defined as
\[
\Phi_{G} = \frac{\displaystyle V(Y)-\prod_{i=1}^R \left[ V(Y \vert X=i) \right]^{p_{i\cdot}}}{V(Y)},
\]
where $V(Y)$ is a measure of variation for the marginal distribution of $Y$.
The geoPRV measure can use the same variation as the conventional PRV measure, for example, 
\[
\Phi_{Gf} = \frac{\displaystyle-\sum_{j=1}^C f(p_{\cdot j}) - \prod_{i=1}^R \left[ -\sum_{j=1}^C f \left( \frac{p_{ij}}{p_{i\cdot}} \right) \right]^{p_{i\cdot}}}{\displaystyle-\sum_{j=1}^C f(p_{\cdot j})},
\]
where the variation measure $V(Y) = -\sum_{j=1}^C f(p_{\cdot j})$.
In addition, the following theorem for $\Phi_{Gf}$ holds.

\begin{theorem}
\label{thm:conditions}
The measure $\Phi_{Gf}$ satisfies the following conditions:
\begin{enumerate}[label = (\roman*), ref = = (\roman*)]
\item $\Phi_f \leq \Phi_{Gf}$. \label{TH10} 
\item $\Phi_{Gf}$ must lie between 0 and 1. \label{TH11}  
\item $\Phi_{Gf}=0$ is equivalent to independence of $X$ and $Y$. \label{TH12}
\item $\Phi_{Gf}=1$ is equivalent to $\prod_{i=1}^R \left[ V(Y \vert X=i) \right]^{p_{i\cdot}}=0$, i.e., for at least one $s$, there exists $t$ such that $p_{st}\neq0$ and $p_{sj}=0$ for every $j$ with $j\neq t$.\label{TH13}
\end{enumerate}
\end{theorem}
\begin{theorem}
\label{thm:invariant}
The value of $\Phi_{Gf}$ is invariant to permutations of row and column categories. 
\end{theorem}
\noindent 
For proof of Theorem \ref{thm:conditions} and Theorem \ref{thm:invariant}, see Appendix \ref{app:conditions} and Appendix \ref{app:invariant}, respectively.
The geoPRV measure differs from the conventional PRV measure in that $\Phi_{Gf}=1$ when there exists $i$ such that $p_{ij}=p_{i\cdot}\neq 0$.
Another important feature of the geoPRV measures is that it takes higher or equal values than the conventional PRV measures, allowing for a stronger representation of row and column relationships.

A property of the geoPRV measure is that the larger the value of $\Phi_{G}$, the stronger the association between the response variable $Y$ and the explanatory variable $X$.
In other words, the larger the value of $\Phi_{G}$, the more accurately you can predict the $Y$ category if you know the $X$ category than if you do not.
In contrast, 
if the value of $\Phi_{G}$ is 0, the $Y$ category is not affected by the $X$ category at all.

\section{Approximate Confidence Interval for the Measure}\label{sec3}
Since the measure $\Phi_G$ is unknown, we derived a confidence interval of $\Phi_G$.
Let $n_{ij}$ denote the frequency for a cell ($i,j$), and $n=\sum_{i=1}^R\sum_{j=1}^C n_{ij}$ ($i=1,2,\ldots,R;~j=1,2,\ldots,C$).
Assume that the observed frequencies $\{ n_{ij}\}$ have a multinomial distribution, we consider an approximate standard error and large-sample confidence interval for $\Phi_G$ using the delta method (\citealp{bishop2007discrete}, and Appendix C in \citealp{agresti2010analysis}).

\begin{theorem}
\label{thm:dlm}
Let $\widehat{\Phi}_{Gf}$ denote a plug-in estimator of $\Phi_{Gf}$.
$\sqrt{n}( \widehat{\Phi}_{Gf}-\Phi_{Gf} )$ converges in distribution to a normal distribution with mean zero and variance $\sigma^2 [ \Phi_{Gf} ]$, where
\begin{equation*}
\sigma^2[\Phi_{Gf}] = \left(\delta^{(f)}\right)^2 \left[ \sum_{i=1}^R\sum_{j=1}^Cp_{ij}(\Delta_{ij}^{(f)})^2 - \left(\sum_{i=1}^R\sum_{j=1}^C p_{ij}\Delta_{ij}^{(f)} \right)^2 \right],
\end{equation*}
with
\begin{eqnarray*}
\delta^{(f)} &=& 
\frac{\displaystyle\prod_{s=1}^R \left[ -\sum_{t=1}^C f \left( \frac{p_{st}}{p_{s\cdot}} \right)  \right]^{p_{s\cdot}}}{\displaystyle\left( \sum_{t=1}^C f(p_{\cdot t}) \right)^2},\\
\Delta_{ij}^{(f)} &=& 
f'(p_{\cdot j}) -\varepsilon_{ij}^{(f)}\sum_{t=1}^C f(p_{\cdot t}),\\
\varepsilon_{ij}^{(f)}	&=& 
\log \left[ -\sum_{t=1}^C f \left( \frac{p_{it}}{p_{i\cdot}} \right) \right] + \frac{\displaystyle\sum_{t=1}^C \left\{ -\frac{p_{it}}{p_{i\cdot}} f' \left( \frac{p_{it}}{p_{i\cdot}} \right) \right\} + f' \left( \frac{p_{ij}}{p_{i\cdot}} \right)}{\displaystyle\sum_{t=1}^C f' \left( \frac{p_{it}}{p_{i\cdot}} \right)},
\end{eqnarray*}
and $f'(x)$ is the derivative of function $f(x)$ by $x$.
\end{theorem}

\noindent
The proof of Theorem \ref{thm:dlm} is given in Appendix \ref{app:dlm}.

Let $\widehat{\sigma}^2 \left[ \Phi_{Gf} \right]$ denote a plug-in estimator of $\sigma^2 \left[ \Phi_{Gf} \right]$.
From Theorem \ref{thm:dlm}, since $\widehat{\sigma} \left[ \Phi_{Gf} \right]$ is a consistent estimator of $\sigma \left[ \Phi_{Gf} \right]$, $\widehat{\sigma} \left[ \Phi_{Gf} \right] / \sqrt{n}$ is an estimated standard error for $\widehat{\Phi}_{Gf}$, and $\widehat{\Phi}_{Gf} \pm z_{\alpha/2} \widehat{\sigma} \left[ \Phi_{Gf} \right] / \sqrt{n}$ is an approximate $100(1-\alpha)\%$ confidence limit for $\Phi_{Gf}$, where $z_{\alpha}$ is the upper two-sided normal distribution percentile at level $\alpha$.

\section{Numerical Experiments}\label{sec4}
In this section, we confirmed the performance of geoPRV measure $\Phi_{Gf}$, and the difference between $\Phi_{Gf}$ and the conventional PRV measure $\Phi_f$ proposed by \cite{momozaki2022extension}.
We use $\Phi_f$ and $\Phi_{Gf}$, which have the variation measure $V(Y)=-\sum_{j=1}^Cf(p_{\cdot j})$. 
In addition to applying $f(x)=\left( x^{\lambda+1} - x \right)/\lambda$ for $\lambda>-1$ and $g(x)=(x - 1)^2/(\omega x + 1 - \omega) - (x - 1)/(1 - \omega)$ for $0 \leq \omega < 1$ (see, \citealp{ichimori2013inequalities}), the former is expressed as $\Phi_f^{(\lambda)}$ and $\Phi_{Gf}^{(\lambda)}$, while the latter is expressed as $\Phi_g^{(\omega)}$ and $\Phi_{Gg}^{(\omega)}$.
For the tuning parameters, set $\lambda=0,~0.5,~1.0$ and $\omega=0,~0.5,~0.9$.

\subsection*{Artificial data 1}
Consider the artificial data in Table \ref{data:simulation_2}.
These are data to clearly show the difference in characteristics between conventional PRV measures and the geoPRV measure.
Table \ref{data:simulation_2}c shows the case where the explanatory variable in the first row has a complete association structure with the response variable in the third column.
On the other hand, Table \ref{data:simulation_2}a and Table \ref{data:simulation_2}b show the case where the explanatory variable in the first row has a weak or slightly strong association structure to the response variable, respectively.

\begin{table}
\centering
\caption{The $3 \times 3$ probability tables, which have a (a) weak (b) slightly strong, and (c) complete association structure in the first row.}
\label{data:simulation_2}
\centering
\begin{tabular}{ccccc}
\hline
 & (1) & (2) & (3) &Total \\ \hline
(a) \\                      
(1) & 0.005   & 0.125  & 0.370  & 0.500  \\
(2) & 0.030   & 0.050  & 0.120  & 0.200  \\
(3) & 0.045   & 0.075  & 0.180 & 0.300  \\ 
Total & 0.080 & 0.250 & 0.670 & 1.000  \\ \hline
(b) \\
(1) & 0.005   & 0.025  & 0.470  & 0.500  \\
(2) & 0.030   & 0.050  & 0.120  & 0.200  \\
(3) & 0.045   & 0.075  & 0.180 & 0.300  \\ 
Total & 0.080 & 0.150 & 0.770 & 1.000  \\ \hline
(c) \\
(1) & 0.000   & 0.000  & 0.500  & 0.500  \\
(2) & 0.030   & 0.050  & 0.120  & 0.200  \\
(3) & 0.045   & 0.075  & 0.180 & 0.300  \\ 
Total & 0.075 & 0.125 & 0.800 & 1.000  \\ \hline
\end{tabular}
\end{table}

The values of $\Phi_f^{(\lambda)}$ and $\Phi_{Gf}^{(\lambda)}$ are provided in Table \ref{result:simulation_2}a and Table \ref{result:simulation_2}b, respectively.
For instance, Table \ref{result:simulation_2}a shows that when Table \ref{data:simulation_2}c is parsed the measure $\Phi_f^{(\lambda)}=0.2628,~0.1990,~0.1784$ for each $\lambda$ and does not capture the complete association structure of the first row.
In contrast, $\Phi_{Gf}^{(\lambda)} = 1$ in all $\lambda$, allowing us to identify the local complete association structure.
Similarly, consider the results of the $\Phi_{Gf}^{(\lambda)} $ and $\Phi_{f}^{(\lambda)}$ in any $\lambda$ from Table \ref{data:simulation_2}a to Table \ref{data:simulation_2}c.
As can be seen from these results, the simulation also shows that $\Phi_{Gf}^{(\lambda)}$ changes significantly by capturing partially related structures compared to $\Phi_{f}^{(\lambda)}$.

\begin{table}[!ht]
\centering
\caption{The value of $\Phi_{f}^{(\lambda)}$ and $\Phi_{Gf}^{(\lambda)}$, applied to Table \ref{data:simulation_2}}
\label{result:simulation_2}
(a) The values of $\Phi_{f}^{(\lambda)}$\\
\begin{tabular}{cccc}
\hline
$\lambda$   & Table \ref{data:simulation_2}a  & Table \ref{data:simulation_2}b & Table \ref{data:simulation_2}c \\ \hline
0.0  & 0.0495 & 0.1285 & 0.2628  \\
0.5  & 0.0302 & 0.1156 & 0.1990 \\
1.0  & 0.0203 & 0.1105 & 0.1784 \\ \hline
\end{tabular}
\end{table}

\begin{table}[!ht]
\centering
(b) The values of  $\Phi_{Gf}^{(\lambda)}$ \\
\begin{tabular}{cccc}
\hline
$\lambda$   & Table \ref{data:simulation_2}a  & Table \ref{data:simulation_2}b & Table \ref{data:simulation_2}c \\ \hline
0.0  & 0.0701 & 0.2765 & 1.0000 \\
0.5  & 0.0487 & 0.3126 & 1.0000 \\
1.0  & 0.0354 & 0.3221 & 1.0000 \\ \hline
\end{tabular}
\end{table}

\subsection*{Artificial data 2}
Consider the artificial data in Table \ref{data:simulation}.
These data are intended to examine the value of the geoPRV measure $\Phi_{Gf}$ as the association of the entire contingency table changes.
Therefore, we obtained data suitable for the survey by converting the bivariate normal distribution with means $\mu_1 = \mu_2 = 0$ and variances $\sigma^2_1 = \sigma^2_2 = 1$, in which the correlation coefficient was changed from $0$ to $1$ by $0.2$, into the $4 \times 4$ contingency tables with equal-interval frequency.
From Theorem \ref{thm:invariant} and the properties of the PRV measures, when the absolute values of the correlation coefficients are the same, i.e., when the rows of the contingency table are simply swapped, the values are equal, so the results for the negative correlation coefficient case are omitted.

\begin{sidewaystable}
\centering
\caption{The $4 \times 4$ probability tables, formed by using three cutpoints for each variable at $z_{0.25}, z_{0.50}, z_{0.75}$ from a bivariate normal distribution with the conditions $\mu_1 = \mu_2 = 0$, $\sigma^2_1 = \sigma^2_2 = 1$, and $\rho$ increasing by 0.2 from $0$ to $1$.}
\label{data:simulation}
\centering
\scalebox{1}{
\begin{tabular}{ccccccccccccc}
\\
\hline
 & (1) & (2) & (3) & (4) &Total &  & (1) & (2) & (3) & (4) &Total \\ \hline
& &\multicolumn{2}{c}{$\rho = 1.0$ } & & &                          & &\multicolumn{2}{c}{$\rho = 0.4$ } & &\\
(1) & 0.2500 & 0.0000 & 0.0000 & 0.0000 & 0.2500 &         (1) & 0.1072 & 0.0692 & 0.0477 & 0.0258 & 0.2500 \\
(2) & 0.0000 & 0.2500 & 0.0000 & 0.0000 & 0.2500 &         (2) & 0.0692 & 0.0698 & 0.0632 & 0.0477 & 0.2500  \\
(3) & 0.0000 & 0.0000 & 0.2500 & 0.0000 & 0.2500 &         (3) & 0.0477 & 0.0632 & 0.0698 & 0.0692 & 0.2500 \\
(4) & 0.0000 & 0.0000 & 0.0000 & 0.2500 & 0.2500 &         (4) & 0.0258 & 0.0477 & 0.0692 & 0.1072 & 0.2500 \\
Total & 0.2500 & 0.2500 & 0.2500  & 0.2500 & 1.0000 &     Total & 0.2500 & 0.2500 & 0.2500 & 0.2500 & 1.0000 \\
& &\multicolumn{2}{c}{$\rho = 0.8$ } & & &                          & &\multicolumn{2}{c}{$\rho = 0.2$ } & &\\
(1) & 0.1691 & 0.0629 & 0.0164 & 0.0016 & 0.2500 &         (1) & 0.0837 & 0.0668 & 0.0563 & 0.0432 & 0.2500 \\
(2) & 0.0629 & 0.1027 & 0.0680 & 0.0164 & 0.2500 &         (2) & 0.0668 & 0.0648 & 0.0621 & 0.0563 & 0.2500 \\
(3) & 0.0164 & 0.0680 & 0.1027 & 0.0629 & 0.2500 &         (3) & 0.0563 & 0.0621 & 0.0648 & 0.0668 & 0.2500 \\
(4) & 0.0016 & 0.0164 & 0.0629 & 0.1691 & 0.2500 &         (4) & 0.0432 & 0.0563 & 0.0668 & 0.0837 & 0.2500 \\
Total & 0.2500 & 0.2500 & 0.2500 & 0.2500 & 1.0000 &      Total & 0.2500 & 0.2500 & 0.2500 & 0.2500 & 1.0000 \\
& &\multicolumn{2}{c}{$\rho = 0.6$ } & & &                          & &\multicolumn{2}{c}{$\rho = 0$ } & &\\
(1) & 0.1345 & 0.0691 & 0.0353 & 0.0111 & 0.2500 &         (1) & 0.0625 & 0.0625 & 0.0625 & 0.0625 & 0.2500 \\
(2) & 0.0691 & 0.0797 & 0.0659 & 0.0353 & 0.2500 &         (2) & 0.0625 & 0.0625 & 0.0625 & 0.0625 & 0.2500 \\
(3) & 0.0353 & 0.0659 & 0.0797 & 0.0691 & 0.2500 &         (3) & 0.0625 & 0.0625 & 0.0625 & 0.0625 & 0.2500 \\
(4) & 0.0111 & 0.0353 & 0.0691 & 0.1345 & 0.2500 &         (4) & 0.0625 & 0.0625 & 0.0625 & 0.0625 & 0.2500 \\
Total & 0.2500 & 0.2500 & 0.2500 & 0.2500 & 1.0000 &      Total & 0.2500 & 0.2500 & 0.2500 & 0.2500 & 1.0000 \\ \hline
\end{tabular}
}
\end{sidewaystable} 

Table \ref{result:simulation1} shows the value of $\Phi_f^{(\lambda)}$ and $\Phi_{Gf}^{(\lambda)}$ for each value of $\rho$, respectively.
We observe that the values of $\Phi_f^{(\lambda)}$ and $\Phi_{Gf}^{(\lambda)}$ increase as the absolute value of the $\rho$ increases.
Besides, $\rho = 0$ if and only if the measures show that it is independent of the table, and $\rho = 1.0$ if and only if the measures confirm that there is a structure of all (or partially) complete association.
Also, if there is a relationship only to the entire contingency table, the values of $\Phi_{Gf}^{(\lambda)}$ are found to be larger than the values of $\Phi_{f}^{(\lambda)}$ by Theorem \ref{thm:conditions}, but the differences are small.

\begin{table}[!ht]
\centering
\caption{The values of $\Phi_f^{(\lambda)}$ and $\Phi_{Gf}^{(\lambda)}$ for each $\rho$}
\label{result:simulation1}
(a) The values of $\Phi_f^{(\lambda)}$ \\
\begin{tabular}{ccccccc} \hline
$\lambda$ & $\rho=0.0$ & $\rho=0.2$ & $\rho=0.4$ & $\rho=0.6$ & $\rho=0.8$ & $\rho=1.0$ \\ \hline
0.0 & 0.0000 & 0.0109 & 0.0461 & 0.1159 & 0.2541 & 1.0000 \\
0.5 & 0.0000 & 0.0113 & 0.0471 & 0.1161 & 0.2479 & 1.0000 \\
1.0 & 0.0000 & 0.0100 & 0.0419 & 0.1035 & 0.2236 & 1.0000 \\ \hline
\end{tabular}
\end{table}

\begin{table}[!ht]
\centering
(b) The values of $\Phi_{Gf}^{(\lambda)}$ \\
\begin{tabular}{ccccccc} \hline
$\lambda$ & $\rho=0.0$ & $\rho=0.2$ & $\rho=0.4$ & $\rho=0.6$ & $\rho=0.8$ & $\rho=1.0$ \\ \hline
0.0 & 0.0000 & 0.0109 & 0.0469 & 0.1203 & 0.2699 & 1.0000 \\
0.5 & 0.0000 & 0.0113 & 0.0479 & 0.1205 & 0.2634 & 1.0000 \\
1.0 & 0.0000 & 0.0100 & 0.0425 & 0.1071 & 0.2369 & 1.0000 \\ \hline
\end{tabular}
\end{table}

\subsection*{Actual data 1}
Consider the case where the PRV measure is adapted to the data in Table \ref{data:classical}, a survey of cannabis use among students conducted at the University of Ioannina (Greece) in 1995 and published in \cite{marselos1997epidemiological}.
The students’ frequency of alcohol consumption is measured on a four-level scale ranging from at most once per month up to more frequently than twice per week while their trial of cannabis through a three-level variable (never tried–tried once or twice–more often). 
We can see the partial bias of the frequency for the first and second rows in the data.

\begin{table}[!ht]
\centering
\caption{Students’ survey about cannabis use at the University of Ioannina}
\label{data:classical}
\begin{tabular}{crrrr}
\hline
\multirow{2}{*}{} & \multicolumn{4}{c}{I tried cannabis $\dots$} \\ \cline{2-5} 
Alcohol consumption & Never & Once or twice & More often & Total \\ \hline
At most once/month	& 204 & 6 & 1 & 211 \\
Twice/month & 211 & 13 & 5 & 229 \\
Twice/week & 357 & 44 & 38 & 439 \\ 
More often & 92 & 34 & 49 & 175 \\ \hline
Total & 864 & 97 & 93 & 1054 \\ \hline
\end{tabular}
\end{table}

The estimates of $\Phi_f^{(\lambda)}$ and $\Phi_{Gf}^{(\lambda)}$ are provided in Table \ref{result:classical_f}a and Table \ref{result:classical_f}b, respectively.
For instance, when $\lambda=1$, the measure $\widehat{\Phi}_f^{(1)}=0.1034$ for Table \ref{result:classical_f}a, and $\widehat{\Phi}_{Gf}^{(1)}=0.2992$ for Table \ref{result:classical_f}b.
$\widehat{\Phi}_f^{(1)}$ shows that the average condition variation of trying cannabis is $10.34\%$ smaller than the marginal variation, and similarly $\widehat{\Phi}_{Gf}^{(1)}$ shows that the average condition variation of trying cannabis is $29.92\%$ smaller.
Based on the results of these values, the following can be interpreted from Table \ref{data:classical}:
\begin{itemize}
\item[(1)] There is a strong association overall between alcohol consumption and cannabis use experience associated.
\item[(2)] There are fairly strong associations between some alcohol consumption and cannabis use experience.
\end{itemize}
These interpretations seem to be intuitive when looking at Table \ref{data:classical}.
However, by analyzing using the measures, we have been able to present an objective interpretation numerically and to show how strongly associated structures are in the contingency table.


\begin{table}[!ht]
\centering
\caption{Estimate of $\Phi_{f}^{(\lambda)}$ and $\Phi_{Gf}^{(\lambda)}$, estimated approximate standard error for $\widehat{\Phi}_{f}^{(\lambda)}$ and $\widehat{\Phi}_{Gf}^{(\lambda)}$, approximate $95\%$ confidence interval for $\Phi_{f}^{(\lambda)}$ and $\Phi_{Gf}^{(\lambda)}$.}
\label{result:classical_f}
(a) $\Phi_{f}^{(\lambda)}$ for Table \ref{data:classical}\\
\begin{tabular}{cccc}
\hline
$\lambda$ & Estimated measure & Standard error & Confidence interval \\ \hline
0.0 & 0.1215 & 0.0175 & (0.0872, 0.1557) \\
0.5 & 0.1090 & 0.0172 & (0.0752, 0.1428) \\
1.0 & 0.1034 & 0.0174 & (0.0693, 0.1376) \\ \hline
\end{tabular}
\end{table}

\begin{table}[!ht]
\centering
(b) $\Phi_{Gf}^{(\lambda)}$ for Table \ref{data:classical}\\
\begin{tabular}{cccc}
\hline
$\lambda$	& Estimated measure & Standard error & Confidence interval \\ \hline
0.0 & 0.2601 & 0.0439 & (0.1741, 0.3461) \\
0.5 & 0.2922 & 0.0488 & (0.1965, 0.3879) \\
1.0 & 0.2992 & 0.0502 & (0.2007, 0.3976) \\ \hline
\end{tabular}
\end{table}

\subsection*{Actual data 2}
By analyzing multiple contingency tables using the measures, it is possible to numerically  determine how much difference there are between the associations of the contingency tables.
Therefore, consider the data in Table \ref{data:NativeClass} are taken from \cite{hashimoto1999gendai}. 
These data describe the cross-classifications of the father's and son's occupational status categories in Japan which were examined in 1975 and 1985. 
In addition, we can consider the father's states as an explanatory variable and the son's states as an response variable, since the father's occupational status categories seem to have an influence on the son's.
The analysis of Table \ref{data:NativeClass} aims to show what differences there are in the associations of occupational status categories for fathers and sons in 1975 and 1985.

\begin{table}[!ht]
\centering
\caption{Occupational status for Japanese father-son pairs}
\label{data:NativeClass}
(a) Examined in 1975 
\begin{tabular}{crrrrr}
\hline
& \multicolumn{4}{c}{Son's status} & \\ \cline{2-5} 
Father's status & Capitalist & New middle& Working& Old middle& Total \\ \hline
Capitalist &29& 43& 25& 35 & 132\\
New middle &23& 159& 89& 52 & 323 \\
Working    &11& 69& 184& 44& 308 \\
Old middle &84& 323& 525& 613& 1545 \\ \hline
Total & 147 & 594& 823 & 744 & 2308 \\ \hline
\end{tabular}
\end{table}

\begin{table}[!ht]
\centering
(b) Examined in 1985 
\begin{tabular}{crrrrr}
\hline
& \multicolumn{4}{c}{Son's status} & \\ \cline{2-5} 
Father's status & Capitalist & New middle& Working& Old middle& Total \\ \hline
Capitalist & 46 &59& 34& 42 & 181\\
New middle& 20& 193& 79& 31 & 323 \\
Working & 9& 122& 202& 48& 381 \\
Old middle &47 & 270& 412& 380& 1109 \\ \hline
Total & 122 & 644& 727 & 501 & 1994 \\ \hline
\end{tabular}
\end{table}

Table \ref{result:NativeClassl_f} and Table \ref{result:NativeClass_Gf} give the estimates of $\Phi_g^{(\omega)}$ and $\Phi_{Gg}^{(\omega)}$, respectively.
Comparing the estimates for each $\omega$ in Table \ref{result:NativeClassl_f} and Table \ref{result:NativeClass_Gf}, we can see that the values for both measures are almost the same.
In addition, comparing Table \ref{result:NativeClassl_f}a and Table \ref{result:NativeClassl_f}b, the estimate is slightly larger in Table \ref{result:NativeClassl_f}b, so it can be assumed that Table \ref{result:NativeClassl_f}b is more related, but there is little difference because all the confidence intervals are covered.
When we also compare \ref{result:NativeClass_Gf}a and Table \ref{result:NativeClass_Gf}b, we can see that \ref{result:NativeClass_Gf}b is larger because the estimate is slightly larger in \ref{result:NativeClass_Gf}b.
However, we can see that the confidence interval does not cover at $\omega = 0.9$.
From the results of these values, the following can be interpreted for Table \ref{data:NativeClass}a and Table \ref{data:NativeClass}b:
\begin{itemize}
\item[(1)] The occupational status categories of fathers and sons in 1975 and 1985 both have weak associations overall, further indicating that individual explanatory variables do not have remarkably associations.
\item[(2)] Although the association of Table \ref{data:NativeClass}b is slightly larger than Table \ref{data:NativeClass}a, the results of the confidence intervals indicate that there is no statistical difference.
\item[(3)] The partial association in Table \ref{data:NativeClass}b is slightly larger than Table \ref{data:NativeClass}a, and the results of confidence intervals indicate that there may be a statistical difference.
\end{itemize}
When there are statistical differences from the results of some confidence intervals, as in (3), it is affected by differences in the characteristics of variation associated with changing the tuning parameters.
In this case, it is difficult to give an interpretation by referring to variation because there was no difference in the variation in the special cases (e.g., $\omega= 0$).
However, when there are differences in variation in special cases, further interpretation can be given by focusing on the characteristics.

\begin{table}[!ht]
\centering
\caption{In the table, the first column indicates the estimate $\widehat{\Phi}_{g}^{(\omega)}$ of $\Phi_{g}^{(\omega)}$, and the second column indicates the estimated approximate standard error for $\widehat{\Phi}_{g}^{(\omega)}$, and the final column indicates approximate $95\%$ confidence interval for $\Phi_{g}^{(\omega)}$.}
\label{result:NativeClassl_f}
(a) For Table \ref{data:NativeClass}a\\
\begin{tabular}{cccc}
\hline
$\omega$ & Estimated measure & Standard error & Confidence interval \\ \hline
0.0& 0.0480& 0.0061& (0.0361, 0.0600) \\
0.5& 0.0547& 0.0067& (0.0416, 0.0678) \\
0.9& 0.0401& 0.0054& (0.0294, 0.0507)    \\ \hline
\end{tabular}
\end{table}

\begin{table}[!ht]
\centering
(b) For Table \ref{data:NativeClass}b\\
\begin{tabular}{cccc}
\hline
$\omega$ & Estimated measure & Standard error & Confidence interval \\ \hline
0.0& 0.0598& 0.0071& (0.0459, 0.0736) \\
0.5& 0.0709& 0.0079& (0.0553, 0.0864) \\
0.9& 0.0665& 0.0081& (0.0506, 0.0823)    \\ \hline
\end{tabular}
\end{table}

\begin{table}[!ht]
\centering
\caption{Estimate of $\Phi_{Gg}^{(\omega)}$, estimated approximate standard error for $\widehat{\Phi}_{Gg}^{(\omega)}$, approximate $95\%$ confidence interval for $\Phi_{Gg}^{(\omega)}$.}
\label{result:NativeClass_Gf}
(a) For Table \ref{data:NativeClass}a\\
\begin{tabular}{cccc}
\hline
$\omega$ & Estimated measure & Standard error & Confidence interval \\ \hline
0.0& 0.0499& 0.0066& (0.0371, 0.0628) \\
0.5& 0.0571& 0.0072& (0.0431, 0.0712) \\
0.9& 0.0416& 0.0057& (0.0304, 0.0528)    \\ \hline
\end{tabular}
\end{table}

\begin{table}[!ht]
\centering
(b) For Table \ref{data:NativeClass}b\\
\begin{tabular}{cccc}
\hline
$\omega$ & Estimated measure & Standard error & Confidence interval \\ \hline
0.0& 0.0630& 0.0077& (0.0478, 0.0782) \\
0.5& 0.0752& 0.0086& (0.0583, 0.0922) \\
0.9& 0.0695& 0.0084& (0.0530, 0.0860)    \\ \hline
\end{tabular}
\end{table}

\section{Conclusion}\label{sec5}
In this paper, we proposed a geometric mean type of PRV (geoPRV) measure that uses variation composed of geometric mean and arbitrary functions that satisfy certain conditions.
We showed that the proposed measure has the following three properties that are suitable for examining the degree of association, which satisfies the conventional measures: (i) The measure increases monotonically as the degree of association increases; (ii) The value is 0 when there is a structure of null association, and (iii) The value is 1 when there is a complete structure of association.
Furthermore, by using geometric means, the geoPRV measure can capture the association to the response variables for individual explanatory variables that could not be investigated by the existing PRV measures.
Analyses using the existing PRV measures and the geoPRV measure simultaneously will be able to examine the association of the entire contingency table and the partial association.
Also, the geoPRV measure can be analyzed using variations with various characteristics by providing functions and tuning parameters that satisfy the conditions, such as the measure $\Phi_{f}$.
Therefore, analysis using the geoPRV measure together can lead to a deeper understanding of the data and provide further interpretation.
While various measures of contingency tables have been proposed, there have been several studies in recent years that have conducted analyses using the Goodman-Kraskal's PRV measure (e.g. \citealp{gea2023biological, iordache2022climate}).
We believe that the new PRV measure in this paper, when examined and compared together with the existing Goodman-Kraskal's PRV measure, may provide a new perspective that pays attention to the association of individual explanatory variables, including the association of the entire contingency table.

\appendix
\section{Proof of Theorem \ref{thm:conditions}}
\label{app:conditions}
\begin{proof}
\begin{enumerate}[label = (\roman*)]
\item Let $\phi$ denote a numerator of a fraction
\[
\Phi_{Gf}-\Phi_f
= \frac{\displaystyle\sum_{i=1}^R p_{i\cdot} V(Y \vert X=i)-\prod_{i=1}^R\left[ V(Y \vert X=i)  \right]^{p_{i\cdot}}}{\displaystyle-\sum_{j=1}^Cf(p_{\cdot j})}.
\]
If there exists $i$ such that $V(Y \vert X=i)=0$, $\phi\geq0$ is easily verified, i.e., $\Phi_f\leq\Phi_{Gf}$ is established.
Moreover, consider cases other than this one.
Assume that $f(x)=-\log x$ which is convex function since $f''(x)=1/x^2>0$ where $f''(x)$ is the second derivative of function $f(x)$ by $x$.
From Jensen's inequality,
\begin{eqnarray*}
&& \sum_{i=1}^R p_{i\cdot}[-\log V(Y \vert X=i)]\geq-\log \left[ \sum_{i=1}^R p_{i\cdot}V(Y \vert X=i) \right]\\
&\Longleftrightarrow& \sum_{i=1}^R \log \left[ V(Y \vert X=i) \right]^{p_{i\cdot}} \leq \log \left[ \sum_{i=1}^R p_{i\cdot}V(Y \vert X=i) \right] \\
&\Longleftrightarrow& \log \prod_{i=1}^R \left[ V(Y \vert X=i) \right]^{p_{i\cdot}} \leq \log \left[ \sum_{i=1}^R p_{i\cdot}V(Y \vert X=i) \right] \\
&\Longleftrightarrow& \prod_{i=1}^R \left[ V(Y \vert X=i) \right]^{p_{i\cdot}} \leq \left[ \sum_{i=1}^R p_{i\cdot}V(Y \vert X=i) \right]
\end{eqnarray*}
where $p_{i\cdot}\geq0$, $\sum_{i=1}^R p_{i\cdot}=1$.
Therefore, $\phi\geq0$, i.e., $\Phi_f\leq\Phi_{Gf}$ holds.
\item The inequality $0\leq\Phi_f\leq1$ is already proven by Momozaki {\it et al.}, and $\Phi_f\leq\Phi_{Gf}$ holds as proved above.
Hence, $\Phi_{Gf}\geq0$ holds since $0\leq\Phi_f\leq\Phi_{Gf}$.
In addition, since $\prod_{i=1}^R \left[ V(Y \vert X=i) \right]^{p_{i\cdot}}\geq0$, we obtain $\Phi_G\leq1$.
Thus, $0\leq\Phi_{Gf}\leq1$ holds.
\item Since $0\leq\Phi_f\leq\Phi_{Gf}$, if $\Phi_{Gf}=0$ then $\Phi_f=0$.
Hence, since $p_{ij}=p_{i\cdot}p_{\cdot j}$ holds for $\Phi_f=0$ (Momozaki {\it et al.}), $p_{ij}=p_{i\cdot}p_{\cdot j}$ holds for $\Phi_{Gf}=0$.
Thus, $\Phi_{Gf}=0\Longrightarrow p_{ij}=p_{i\cdot}p_{\cdot j}$ holds.
Moreover, $\Phi_{Gf}=0\Longleftarrow p_{ij}=p_{i\cdot}p_{\cdot j}$ can be easily checked.
\item If $\Phi_{Gf}=1$ then $\prod_{i=1}^R \left[ V(Y \vert X=i) \right]^{p_{i\cdot}}=0$, i.e., for some $s$, $V(Y \vert X=s)=-\sum_{j=1}^Cf\left( \frac{p_{ij}}{p_{i\cdot}} \right)=0$ ($s=1,2,\ldots,R$).
Thus, there exists $i$ such that $p_{ij}\neq0$ and $p_{ik}=0$ ($k\neq j$).
\end{enumerate}
\end{proof}

\section{Proof of Theorem \ref{thm:invariant}}
\label{app:invariant}
\begin{proof}
Since the first terms in the denominator and numerator of $\Phi_{Gf}$ do not depend on the row category, we focus on the second term in the numerator.
This term is
\begin{equation*}
    \prod_{i=1}^R \left[ - \sum _{j=1} ^C f 
    \left( \frac{p_{ij}}{p_{i\cdot}} \right) \right]^{p_{i\cdot}} 
    = \prod_{i=1}^R \left[ -f \left(\frac{p_{i1}}{p_{i\cdot}} \right) - \cdots - f \left(\frac{p_{iC}}{p_{i\cdot}}\right) \right]^{p_{i\cdot}},
\end{equation*}
and the values are invariant to the reordering of the sums.
Namely, the value of $\Phi_{Gf}$ is invariant with respect to the permutation of row categories.
Similarly, the value of $\Phi_{Gf}$ is also invariant with respect to the permutation of column categories.
\end{proof}

\section{Proof of Theorem \ref{thm:dlm}}
\label{app:dlm}
\begin{proof}
Let
\[
\bm{n}=(n_{11},n_{12},\ldots,n_{1C},n_{21},\ldots,n_{RC})^\top,
\]
\[
\bm{p}=(p_{11},p_{12},\ldots,p_{1C},p_{21},\ldots,p_{RC})^\top,
\]
$\widehat{\bm{p}}=\bm{n}/n$, and $\bm{a}^\top$ is a transpose of $\bm{a}$.
Then $\sqrt{n}\left( \widehat{\bm{p}} - \bm{p} \right)$ converges in distribution to a normal distribution with mean zero and the covariance matrix $\diag(\bm{p}) - \bm{pp}^\top$, where $\diag(\bm{p})$ is a diagonal matrix with the elements of $\bm{p}$ on the main diagonal (\citealp{bishop2007discrete}).

The Taylor expansion of the function $\widehat{\Phi}_{Gf}$ around $\bm{p}$ is given by
\[
\widehat{\Phi}_{Gf} = \Phi_{Gf} + \left( \frac{\pd \Phi_{Gf}}{\pd \bm{p}^\top} \right) (\widehat{\bm{p}}-\bm{p}) + o_p(n^{-1/2}).
\]
Therefore, since
\[
\sqrt{n}(\widehat{\Phi}_{Gf}-\Phi_{Gf}) = \sqrt{n} \left( \frac{\pd \Phi_{Gf}}{\pd \bm{p}^\top} \right) (\widehat{\bm{p}}-\bm{p}) + o_p(1),
\]
\[
\sqrt{n}(\widehat{\Phi}_{Gf}-\Phi_{Gf}) \overset{d}{\to} N(0,\sigma^2[\Phi_{Gf}]),
\]
where
\[
\sigma^2[\Phi_{Gf}] = \left(\delta^{(f)}\right)^2 \left[ \sum_{i=1}^R\sum_{j=1}^Cp_{ij}(\Delta_{ij}^{(f)})^2 - \left(\sum_{i=1}^R\sum_{j=1}^C p_{ij}\Delta_{ij}^{(f)} \right)^2 \right],
\]
with
\begin{eqnarray*}
\delta^{(f)} &=& 
\frac{\displaystyle\prod_{s=1}^R \left[ -\sum_{t=1}^C f \left( \frac{p_{st}}{p_{s\cdot}} \right)  \right]^{p_{s\cdot}}}{\displaystyle\left( \sum_{t=1}^C f(p_{\cdot t}) \right)^2}\\
\Delta_{ij}^{(f)} &=& 
f'(p_{\cdot j}) -\varepsilon_{ij}^{(f)}\sum_{t=1}^C f(p_{\cdot t}),\\
\varepsilon_{ij}^{(f)}	&=& 
\log \left[ -\sum_{t=1}^C f \left( \frac{p_{it}}{p_{i\cdot}} \right) \right] + \frac{\displaystyle\sum_{t=1}^C \left\{ -\frac{p_{it}}{p_{i\cdot}} f' \left( \frac{p_{it}}{p_{i\cdot}} \right) \right\} + f' \left( \frac{p_{ij}}{p_{i\cdot}} \right)}{\displaystyle\sum_{t=1}^C f' \left( \frac{p_{it}}{p_{i\cdot}} \right)},
\end{eqnarray*}
and $f'(x)$ is the derivative of function $f(x)$ by $x$.
\end{proof}

\bibliographystyle{apalike} 
\bibliography{sn-bibliography2.bib}

\begin{thebibliography}{}

\bibitem[Agresti, 2003]{agresti2003categorical}
Agresti, A. (2003).
\newblock {\em Categorical data analysis}.
\newblock John Wiley \& Sons.

\bibitem[Agresti, 2010]{agresti2010analysis}
Agresti, A. (2010).
\newblock {\em Analysis of ordinal categorical data}, volume 656.
\newblock John Wiley \& Sons.

\bibitem[Bishop et~al., 2007]{bishop2007discrete}
Bishop, Y.~M., Fienberg, S.~E., and Holland, P.~W. (2007).
\newblock {\em Discrete multivariate analysis: theory and practice}.
\newblock Springer Science \& Business Media.

\bibitem[Cram{\'e}r, 1999]{cramer1999mathematical}
Cram{\'e}r, H. (1999).
\newblock {\em Mathematical methods of statistics}, volume~43.
\newblock Princeton university press.

\bibitem[Everitt, 1992]{everitt1992analysis}
Everitt, B.~S. (1992).
\newblock {\em The analysis of contingency tables}.
\newblock CRC Press.

\bibitem[Gea-Izquierdo, 2023]{gea2023biological}
Gea-Izquierdo, E. (2023).
\newblock Biological risk of legionella pneumophila in irrigation systems.
\newblock {\em Revista de Salud P{\'u}blica}, 22:434--439.

\bibitem[Gilula and Haberman, 1986]{gilula1986canonical}
Gilula, Z. and Haberman, S.~J. (1986).
\newblock Canonical analysis of contingency tables by maximum likelihood.
\newblock {\em Journal of the American statistical association},
  81(395):780--788.

\bibitem[Goodman, 1981]{goodman1981association}
Goodman, L.~A. (1981).
\newblock Association models and canonical correlation in the analysis of
  cross-classifications having ordered categories.
\newblock {\em Journal of the American Statistical Association},
  76(374):320--334.

\bibitem[Goodman, 1985]{goodman1985analysis}
Goodman, L.~A. (1985).
\newblock The analysis of cross-classified data having ordered and/or unordered
  categories: Association models, correlation models, and asymmetry models for
  contingency tables with or without missing entries.
\newblock {\em The Annals of Statistics}, 13:10--69.

\bibitem[Goodman and Kruskal, 1954]{doi:10.1080/01621459.1954.10501231}
Goodman, L.~A. and Kruskal, W.~H. (1954).
\newblock Measures of association for cross classifications.
\newblock {\em Journal of the American Statistical Association},
  49(268):732--764.

\bibitem[Hashimoto, 1999]{hashimoto1999gendai}
Hashimoto, K. (1999).
\newblock Gendai nihon no kaikyuu kouzou (class structure in modern japan:
  theory, method and quantitative analysis).
\newblock {\em Toshindo, Tokyo (in Japanese)}.

\bibitem[Ichimori, 2013]{ichimori2013inequalities}
Ichimori, T. (2013).
\newblock On inequalities between $f$-divergence.
\newblock {\em Technical Note, {IPSJ} Journal}, 54(11):2344--2348.

\bibitem[Iordache et~al., 2022]{iordache2022climate}
Iordache, A.~M., Nechita, C., Voica, C., Pluh{\'a}{\v{c}}ek, T., and Schug,
  K.~A. (2022).
\newblock Climate change extreme and seasonal toxic metal occurrence in
  romanian freshwaters in the last two decades—case study and critical
  review.
\newblock {\em NPJ Clean Water}, 5(1):2.

\bibitem[Kateri and Papaioannou, 1994]{kateri1994f}
Kateri, M. and Papaioannou, T. (1994).
\newblock {\em f-divergence Association Models}.
\newblock University of Ioannina.

\bibitem[Marselos et~al., 1997]{marselos1997epidemiological}
Marselos, M., Boutsouris, K., Liapi, H., Malamas, M., Kateri, M., and
  Papaioannou, T. (1997).
\newblock Epidemiological aspects of the use of cannabis among university
  students in greece.
\newblock {\em European Addiction Research}, 3(4):184--191.

\bibitem[Momozaki et~al., 2022]{momozaki2022extension}
Momozaki, T., Wada, Y., Nakagawa, T., and Tomizawa, S. (2022).
\newblock Extension of generalized proportional reduction in variation measure
  for two-way contingency tables.
\newblock {\em Behaviormetrika}, pages 1--14.

\bibitem[Nakagawa et~al., 2020]{nakagawa2020geometric}
Nakagawa, T., Takei, T., Ishii, A., and Tomizawa, S. (2020).
\newblock Geometric mean type measure of marginal homogeneity for square
  contingency tables with ordered categories.
\newblock {\em Journal of Mathematics and Statistics}, 16(1):170--175.

\bibitem[Patil and Taillie, 1982]{patil1982diversity}
Patil, G. and Taillie, C. (1982).
\newblock Diversity as a concept and its measurement.
\newblock {\em Journal of the American statistical Association},
  77(379):548--561.

\bibitem[Rom and Sarkar, 1992]{rom1992generalized}
Rom, D. and Sarkar, S.~K. (1992).
\newblock A generalized model for the analysis of association in ordinal
  contingency tables.
\newblock {\em Journal of statistical planning and inference}, 33(2):205--212.

\bibitem[Saigusa et~al., 2016]{saigusa2016measure}
Saigusa, Y., Tahata, K., and Tomizawa, S. (2016).
\newblock Measure of departure from partial symmetry for square contingency
  tables.
\newblock {\em Journal of Mathematics and Statistics}, 12(3):152--156.

\bibitem[Saigusa et~al., 2019]{saigusa2019measure}
Saigusa, Y., Takami, M., Ishii, A., Nakagawa, T., and Tomizawa, S. (2019).
\newblock Measure for departure from cumulative partial symmetry for square
  contingency tables with ordered categories.
\newblock {\em Journal of Statistics: Advances in Theory and Applications},
  21:53--70.

\bibitem[Tahata, 2022]{tahata2022advances}
Tahata, K. (2022).
\newblock Advances in quasi-symmetry for square contingency tables.
\newblock {\em Symmetry}, 14(5):1051.

\bibitem[Theil, 1970]{theil1970estimation}
Theil, H. (1970).
\newblock On the estimation of relationships involving qualitative variables.
\newblock {\em American Journal of Sociology}, 76(1):103--154.

\bibitem[Tomizawa et~al., 2004]{tomizawa2004generalization}
Tomizawa, S., Miyamoto, N., and Houya, H. (2004).
\newblock Generalization of cramer's coefficient of association for contingency
  tables: theory and methods.
\newblock {\em South African Statistical Journal}, 38(1):1--24.

\bibitem[Tomizawa et~al., 1997]{tomizawa1997generalized}
Tomizawa, S., Seo, T., and Ebi, M. (1997).
\newblock Generalized proportional reduction in variation measure for two-way
  contingency tables.
\newblock {\em Behaviormetrika}, 24(2):193--201.

\bibitem[Tschuprow, 1925]{Tschuprow1926grundbegriffe}
Tschuprow, A. (1925).
\newblock {\em Grundbegriffe und grundprobleme der korrelationstheorie}.
\newblock Leipzig: B.G. Teubner.

\bibitem[Tschuprow, 1939]{Tschuprow1939principles}
Tschuprow, A. (1939).
\newblock {\em Principles of the mathematical theory of correlation}.
\newblock W. Hodge \& Co.

\end{thebibliography}
\end{document}